\documentclass[amsmath,floatfix,twocolumn,superscriptaddress]{revtex4-1}
\usepackage{subfigure}
\usepackage{siunitx}
\usepackage{amssymb}
\usepackage{amsmath}
\usepackage{graphicx}
\usepackage{array}
\usepackage{dcolumn}
\usepackage{psfrag}
\usepackage{bm}
\usepackage{color}
\usepackage{multirow}

\begin{document}

\title{Why Physics Still Matters: Improving Machine Learning Prediction of Material Properties with Phonon-Informed Datasets}

\author{Pol Benítez}
\email{pol.benitez@upc.edu}
    \affiliation{Department of Physics, Universitat Politècnica de Catalunya, 08034 Barcelona, Spain}
    \affiliation{Research Center in Multiscale Science and Engineering, Universitat Politècnica de Catalunya, 
    08019 Barcelona, Spain}

\author{Cibrán López}
    \affiliation{Department of Physics, Universitat Politècnica de Catalunya, 08034 Barcelona, Spain}
    \affiliation{Research Center in Multiscale Science and Engineering, Universitat Politècnica de Catalunya, 
    08019 Barcelona, Spain}

\author{Edgardo Saucedo}
    \affiliation{Research Center in Multiscale Science and Engineering, Universitat Politècnica de Catalunya, 
    08019 Barcelona, Spain}
    \affiliation{Department of Electronic Engineering, Universitat Politècnica de Catalunya, 08034 Barcelona, Spain}

\author{Teruyasu Mizoguchi}
    \affiliation{Institute of Industrial Science, The University of Tokyo, Tokyo 153-8505, Japan}

\author{Claudio Cazorla}
\email{claudio.cazorla@upc.edu}
    \affiliation{Department of Physics, Universitat Politècnica de Catalunya, 08034 Barcelona, Spain}
    \affiliation{Research Center in Multiscale Science and Engineering, Universitat Politècnica de Catalunya, 
    08019 Barcelona, Spain}
    \affiliation{Institució Catalana de Recerca i Estudis Avançats (ICREA), Passeig Lluís Companys 23, 
    		08010 Barcelona, Spain}

\begin{abstract}
\textbf{Abstract.}~Machine learning (ML) methods have become powerful tools for predicting material properties with near first-principles 
	accuracy and vastly reduced computational cost. However, the performance of ML models critically depends on the quality, size, 
	and diversity of the training dataset. In materials science, this dependence is particularly important for learning from 
	low-symmetry atomistic configurations that capture thermal excitations, structural defects, and chemical disorder, features that 
	are ubiquitous in real materials but underrepresented in most datasets. The absence of systematic strategies for generating 
	representative training data may therefore limit the predictive power of ML models in technologically critical fields such as 
	energy conversion and photonics. In this work, we assess the effectiveness of graph neural network (GNN) models trained on two 
	fundamentally different types of datasets: one composed of randomly generated atomic configurations and another 
	constructed using physically informed sampling based on lattice vibrations. As a case study, we address the challenging task of 
	predicting electronic and mechanical properties of a prototypical family of optoelectronic materials under realistic 
	finite-temperature conditions. We find that the phonons-informed model consistently outperforms the randomly trained counterpart, 
	despite relying on fewer data points. Explainability analyses further reveal that high-performing models assign greater weight to 
	chemically meaningful bonds that control property variations, underscoring the importance of physically guided data generation. 
	Overall, this work demonstrates that larger datasets do not necessarily yield better GNN predictive models and introduces a simple 
	and general strategy for efficiently constructing high-quality training data in materials informatics.
\\

{\bf Keywords:} materials informatics, physics-informed machine learning, graph neural networks, model explainability, energy 
	        materials

\end{abstract}

\maketitle

\section{Introduction}
\label{sec:intro}
Over the past decade, machine learning (ML) methods have transformed nearly every field of engineering and natural and social sciences. 
The exponential increase in available data, combined with advances in computational power and algorithmic innovation, has allowed ML models 
to achieve remarkable successes in domains once considered uniquely human or prohibitively complex. From natural language processing, 
exemplified by large language models \cite{zhao2023survey} capable of generating coherent text, reasoning about problems, and accelerating 
scientific discovery, to breakthroughs in biology such as DeepMind's AlphaFold \cite{jumper2021highly}, which revolutionized protein 
structure prediction, ML has demonstrated an unprecedented ability to learn from data and extract patterns beyond human intuition.

In the context of materials science, ML developments have been equally transformative  \cite{malica2025artificial}. The field has 
traditionally relied on a combination of experimental synthesis, characterization, and theoretical modeling to understand and design 
materials. However, these approaches can be time and resources intensive, especially when exploring vast compositional and structural 
spaces. ML offers a complementary paradigm, one in which predictive models can rapidly screen candidate materials \cite{xia2022accelerating}, 
infer hidden correlations between compositions, structure, and properties \cite{liu2024interpretable, karimitari2024accurate}, and  
guide experiments toward promising regions of compositional space \cite{ren2018accelerated, zeni2025generative}. As a result, ML has 
become a central component of the emerging discipline of materials informatics, enabling accelerated discovery and optimization across 
applications such as catalysis, energy storage, and structural alloys.

Yet, despite the growing success of ML techniques in materials discovery, the accuracy and reliability of ML models ultimately 
depend on the quality, diversity, and physical representativeness of the underlying dataset \cite{himanen2019data}. In practice, 
materials datasets are often scarce, noisy, or biased toward well-studied chemical systems, resulting in models that generalize 
poorly when extrapolated to unexplored compositions and structures. Consequently, dataset curation has become as critical 
as model architecture itself, and balancing data quantity with data quality remains a central challenge in materials informatics.

Many ML approaches implicitly assume that enlarging the dataset will systematically improve model accuracy \cite{schmidt2024improving}. 
However, producing large-scale datasets through high-fidelity first-principles simulations can be computationally prohibitive, often 
surpassing the cost of the calculations that these models aim to replace. This difficulty raises a key question: how can one achieve 
reliable predictions from limited high-quality data? Addressing this problem requires careful assessment of model convergence with 
respect to dataset size and an understanding that well-designed and diversified datasets in principle may outperform much larger 
scattered ones \cite{zhang2018strategy,xu2023small}.

Actually, to overcome data scarcity and improve generalization, physics-informed ML has emerged as a promising paradigm 
\cite{karniadakis2021physics,jha2018elemnet}. By embedding fundamental physical principles (e.g., symmetry constraints, conservation laws, 
and thermodynamic relations) into the learning process, such approaches enforce physical consistency, enhance interpretability, and reduce 
the need for extensive data. Incorporating domain knowledge in this way allows ML models to learn not only from data correlations but also 
from the governing laws of nature, bridging the gap between statistical learning and physical insight.

One domain where these considerations are particularly pressing is the study of thermal behaviour of functional materials 
\cite{cardona2004temperature,masuki2022ab}. Thermal lattice vibrations, or phonons, strongly influence electronic and mechanical properties 
by stabilizing specific crystal phases at elevated temperatures \cite{rurali2024giant} or modifying the band structure and band gap of
crystals \cite{monserrat2018electron}. Capturing such effects with first-principles methods typically requires computationally demanding 
finite-temperature simulations, which severely limit the size of available datasets. Consequently, there is a strong motivation to develop 
ML models that can serve as accurate surrogates for these high-cost simulations, provided the training data faithfully represent thermally 
disordered atomic configurations.

In this work, we explore how fundamental properties of technologically relevant materials, specifically, anti-perovskites used in 
photovoltaics, electrochemistry, and catalysis \cite{cano2024novel,antiperov1,antiperov2,antiperov3,antiperov4}, evolve with temperature 
using graph neural networks (GNNs). We systematically compare GNN models trained on two different types of low-symmetry, non-equilibrium 
structures: (i)~randomly disordered atomic configurations that broadly sample the configurational space and (ii)~physics-informed phonon 
displacements that selectively probe the low-energy subspace accessible to ions in crystals. Our results show that models trained on 
phonon-informed datasets achieve higher accuracy and robustness with significantly fewer data points. Explainability analyses further reveal 
that these models assign greater importance to chemically meaningful bonds governing band-gap variations, thereby linking predictive 
performance to physical interpretability. Overall, this study demonstrates that embedding physical knowledge in dataset construction can 
substantially enhance ML performance in materials science and provides a pathway toward more efficient and physically grounded ML-driven 
materials discovery.

\begin{figure*}
    \centering
    \includegraphics[width=1\linewidth]{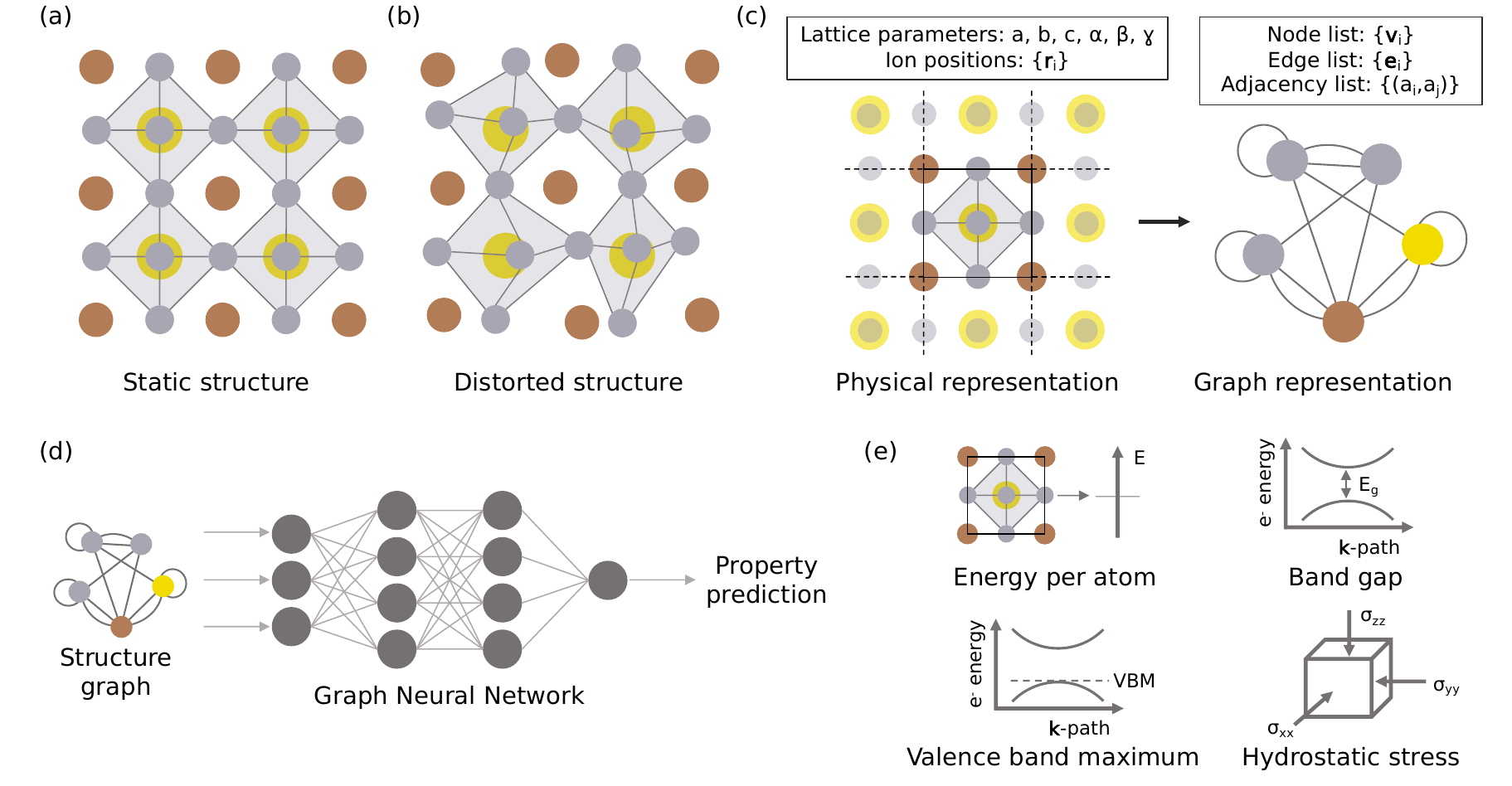}
    \caption{\textbf{Physical and ML aspects in materials informatics covered in this study}. Atomic structure of the reference 
	anti-perovskite system with ions located (a)~at the equilibrium lattices sites and (b)~around the equilibrium lattice sites. 
	(c)~Mapping of a real-space atomic configuration into its graph representation. Crystal periodicity is indicated by dashed 
	lines in the physical representation and self-loops and multiple edges in the graph representation. (d)~Property prediction 
	using GNNs. (e)~The selected quantities for ML prediction: energy per atom, band gap, valence band maximum, and hydrostatic 
	stress.}
    \label{fig1}
\end{figure*}

\section{Results}
\label{sec:results}
GNN models were trained to reproduce the physical behaviour of anti-perovskite materials under realistic $T \neq 0$~K conditions.
The predictive capabilities of the developed GNN frameworks are systematically assessed as follows. First, the performance of the ML 
models trained on a large dataset of non-equilibrium atomic configurations is presented, focusing on relevant electronic and mechanical 
properties. Next, the influence of dataset quality and physical relevance on model accuracy is examined by comparing models trained 
under distinct atomic displacement-generation schemes. Finally, explainability analyses are employed to identify the atomic-scale 
features that most strongly govern the predictive behavior of the developed GNN models, establishing direct physicochemical connections 
between atomic position fluctuations and finite-temperature optoelectronic properties.

\subsection{Dataset and graph generation}
\label{subsec:generalities}
Crystals with the general chemical formula ABX$_3$ typically adopt a centrosymmetric cubic $Pm\overline{3}m$ structure at moderate 
and high temperatures, as exemplified by archetypal materials such as BaTiO$_3$ \cite{smith2008crystal} and Ag$_3$SBr 
\cite{benitez2025crystal}. In this reference crystal structure, BX$_6$ octahedra are formed with B ions located at the octahedral 
centers, X ions at the vertices, and A ions occupying the corners of the cubic unit cell (Fig.~\ref{fig1}a). When A and B species 
are cations and X species are anions, the resulting centrosymmetric phase is known as a perovskite. Conversely, when A and B are 
anions and X are cations, the structure is referred to as an anti-perovskite. 

Anti-perovskite materials are of great technological relevance for energy applications operating at ambient and higher temperatures, 
including solar cells and electrochemical batteries \cite{cano2024novel,antiperov1,antiperov2,antiperov3,antiperov4}. Recent studies 
have revealed that the optoelectronic properties of silver chalcohalide anti-perovskites, Ag$_{3}$XY (X=S,Se; Y=Br,I), may exhibit 
a pronounced dependence on temperature \cite{benitez2025giant,benitez2025band}. In particular, quantities such as the band gap and 
light absorption coefficient may vary by large percentages ($\sim 10$\%) due to temperature-induced volume expansion and local 
symmetry breaking. 

In this work, we generated a comprehensive first-principles dataset for the silver chalcohalide anti-perovskites Ag$_3$SBr, Ag$_3$SI, 
and Ag$_3$SBr$_{x}$I$_{1-x}$. Specifically, the energy per atom, band gap, valence band maximum, and hydrostatic stress were accurately 
calculated using density functional theory (DFT) for a total of $4,500$ non-equilibrium configurations (Sec.~\ref{subsec:performance} 
and Fig.~\ref{fig1}e). These configurations capture the thermal motion of atoms at $T \neq 0$~K and are therefore suitable for training 
ML models aimed at predicting the physical properties of silver chalcohalide anti-perovskites under realistic thermal conditions.

\begin{figure*}
    \centering
    \includegraphics[width=1\linewidth]{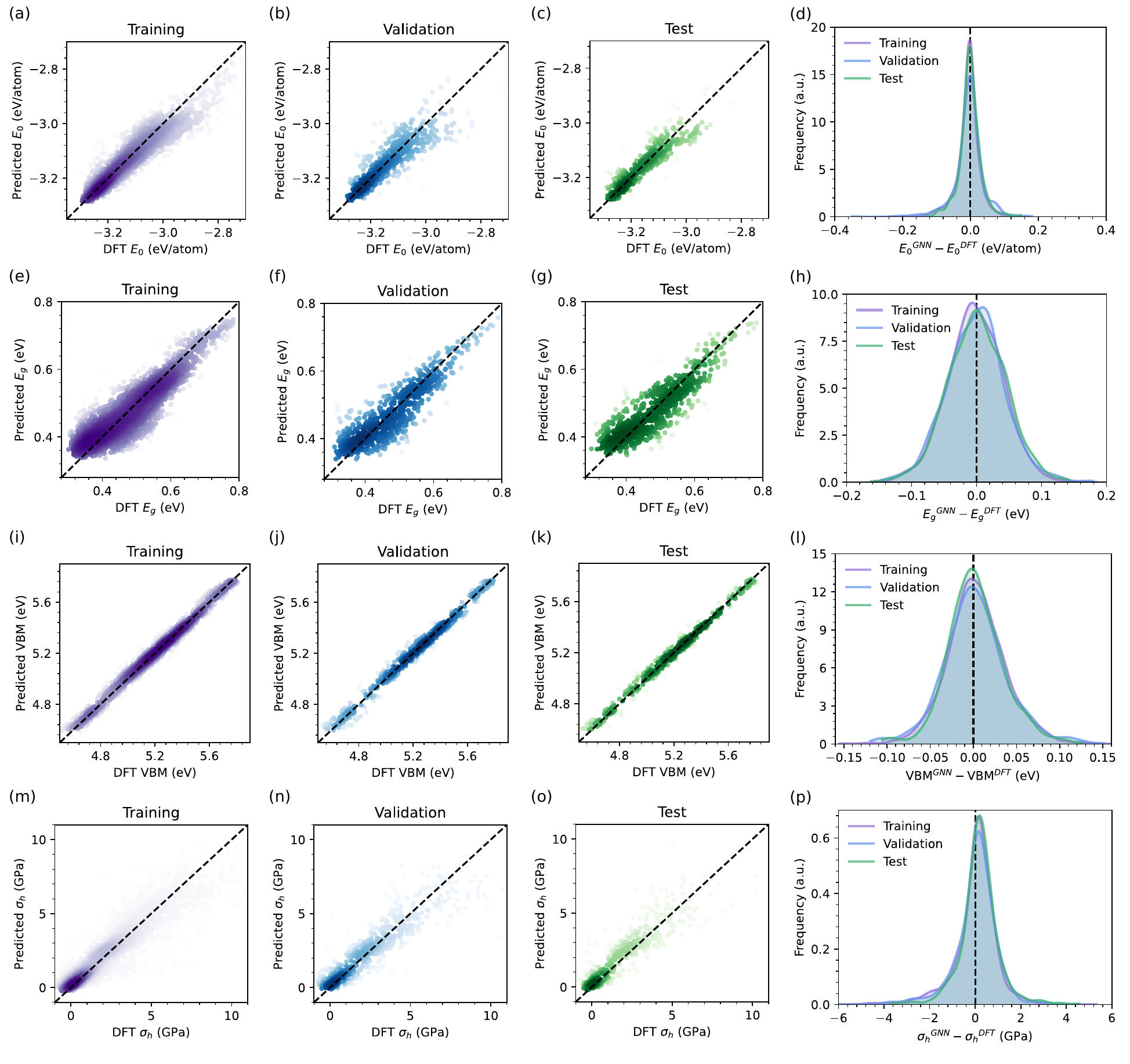}
    \caption{\textbf{Performance of best GNN models}. DFT values versus GNN model predictions for the training, validation,
        and test datasets. Normalized distribution of errors in ML predictions are shown in (d),(h),(l),(p). Results are
        shown for (a--d) energy per atom, (e--h) band gap, (i--l) valence band maximum, and (m--p) hydrostatic stress.}
    \label{fig2}
\end{figure*}

\begin{table*}[!htbp]
\begin{tabular}{c | c c c | c c c | c c c | c c c}
\hline
\hline
% & & & & & & & & & & & & \\
 & \multicolumn{3}{c|}{$E_{0}$ (eV/atom)}  &  \multicolumn{3}{c|}{$E_{g}$ (eV)}  &  \multicolumn{3}{c|}{VBM (eV)}  &  \multicolumn{3}{c}{$\sigma_{h}$ (GPa)}  \\
% & & & & & & & & & & & & \\
\hline
% & & & & & & & & & & & & \\
 & \quad  MAE  \quad   & \quad RMSE  \quad   & \quad $R^2$  \quad & \quad  MAE \quad   & \quad RMSE  \quad   & \quad  $R^2$  \quad  & \quad  MAE  \quad  & \quad  RMSE  \quad   & \quad  $R^2$  \quad &  \quad  MAE  \quad  & \quad RMSE  \quad  & \quad $R^2$ \quad \\
% & & & & & & & & & & & & \\
 \hline
% & & & & & & & & & & & & \\
 Train      &  0.024    &  0.036   &  0.89    &  0.034    &  0.043   &  0.81    &  0.026  & 0.035  & 0.99   & 0.65   &  1.01  &  0.84  \\
 Validation &  0.028    &  0.044   &  0.79    &  0.035    &  0.045   &  0.79    &  0.028  & 0.037  & 0.98   & 0.65   &  0.99  &  0.84  \\
 Test       &  0.022    &  0.032   &  0.85    &  0.035    &  0.044   &  0.79    &  0.024  & 0.033  & 0.99   & 0.60   &  0.88  &  0.82  \\
% & & & & & & & & & & & & \\
 \hline
 \hline
\end{tabular}
\caption{{\bf Best model performance metrics.} MAE, RMSE, and $R^2$ values for the training, validation, and test sets of the best GNNs model
        in predicting energy per atom, band gap, valence band maximum, and hydrostatic stress. MAE and RMSE are reported in the units of the
        corresponding property, while $R^2$ is dimensionless.}
\label{tab:bandgap}
\end{table*}

For crystalline materials, particular attention must be paid to properly preserve translational invariance when mapping atomic structures 
onto graph representations. The strategy that we employ to incorporate crystal periodicity into our GNN models is detailed in the Methods 
section and Supplementary Information (Supplementary Fig.~1), and is schematically illustrated in Fig.~\ref{fig1}c. While a crystal 
structure is defined by its lattice vectors ($a$, $b$, $c$, $\alpha$, $\beta$, $\gamma$) and atomic positions ${\mathbf{r}_i}$, the 
corresponding graph representation is described by the node tensor ${\mathbf{v}_i}$ (atomic features), the edge tensor ${\mathbf{e}_{ij}}$ 
(bonding features), and the adjacency tensor ${(a_i, b_j)}$ (connectivity information). Once a graph is constructed, either for an 
equilibrium configuration or for perturbed atomic positions, it is fed into a GNN model (Fig.~\ref{fig1}d). 

As described in the Methods section and Supplementary Information, for the graph generation we introduced a cutoff radius to determine which 
atomic pairs are connected in the graph, corresponding to the maximum interatomic distance considered for a chemical bond. Supplementary 
Fig.~2 presents the model performance as a function of this cutoff radius. Shorter cutoff values produce simpler graphs and faster training 
but capture fewer interatomic interactions, whereas larger radii increase computational cost without necessarily enhancing predictive 
accuracy. We found that a cutoff radius of $5.5$~\AA~ yields well-converged results, which was adopted throughout this work.

\subsection{GNN model training}
\label{subsec:train}
A comprehensive hyperparameter study was performed to identify the optimal architecture and training parameters for each target property. 
Six distinct model architectures, differing in the number and type of layers, were evaluated (Supplementary Fig.~3). The graph convolution 
layer was implemented following the approach of Morris \textit{et al.} \cite{morris2019weisfeiler}, in which the edge weights, corresponding 
to the Euclidean distances between connected atoms, modulate the message-passing operation. This transformation can be expressed as:
\begin{equation}
    \mathbf{x'}_i = \mathbf{W}_1 \mathbf{x}_i + \mathbf{W}_2 \sum_{j \in \mathcal{N}(i)} e_{j,i}\cdot \mathbf{x}_j,
\end{equation}
where $\mathbf{x}_i$ and $\mathbf{x}'_i$ denote the node feature vectors before and after the convolution layer, respectively, $\mathbf{W}_1$ 
and $\mathbf{W}_2$ are trainable weight matrices, and $e_{j,i}$ is the edge weight between nodes $i$ and $j$. The following hyperparameters 
were explored: ($10^{-2}$, $10^{-3}$, $10^{-4}$, $10^{-5}$, $10^{-6}$) for learning rate, (32, 64, 128, 256) for number of hidden channels, 
and (0.0, 0.2, 0.4, 0.6) for dropout. As a result, a total of $480$ GNN models were trained for each property.

The full dataset comprised $4,500$ configurations generated for anti-perovskite materials (Sec.~\ref{subsec:generalities}) using two 
different schemes for sampling thermal ionic motion, which are discussed in detail in the next subsection. The dataset was then divided 
into 70\% for training, 15\% for validation, and 15\% for testing. To ensure consistent and meaningful comparisons, all models were 
trained on the same training set and evaluated on the same validation set. The best-performing model for each target property was 
subsequently assessed on the independent test set.

Figure~\ref{fig2} presents the performance of the GNN models that best reproduce the DFT reference values across the training, validation, 
and test sets. Results are shown for the energy per atom (Figs.~\ref{fig2}a--c), band gap (Figs.~\ref{fig2}e--g), valence band maximum 
(Figs.~\ref{fig2}i--k), and hydrostatic stress (Figs.~\ref{fig2}m--o). The corresponding distributions of prediction errors are displayed in 
Figs.~\ref{fig2}d,h,l,p. In the scatter plots, the color intensity reflects the point density, with darker colors indicating higher density. 
Overall, the GNN models exhibit strong predictive performance across all datasets and properties, as most points cluster near the diagonal 
dashed line, indicating close agreement with DFT. The error distributions are centered around zero, showing no systematic over- or 
underestimation. Among the studied quantities, the band gap displays the largest prediction dispersion, consistent with previous reports 
\cite{xie2018crystal}, reflecting its higher intrinsic complexity.

Table~I summarizes key performance metrics for the best GNN models, including the mean absolute error (MAE), root mean squared error (RMSE), 
and squared Pearson correlation coefficient ($R^2$). The similarity of results across the training, validation, and test sets confirms the 
absence of overfitting. The MAE values were approximately $0.025$~eV/atom for the energy, $0.035$~eV for both the band gap and valence band 
maximum, and $0.65$~GPa for the hydrostatic stress; these figures are comparable to typical DFT accuracies and to results reported for other 
GNN-based models. The $R^2$ values were around or above $0.80$ in all cases, confirming strong correlations between GNN predictions and DFT 
calculations. Overall, these results demonstrate that GNN models can effectively and efficiently reproduce DFT-level predictions for 
atomically disordered anti-perovskite systems.

Finally, Supplementary Figs.~4--7 summarize the performance of all trained models (excluding those with $R^2 < -1$, which perform worse than 
random guessing) over the validation set. The results are displayed as violin plots, showing the distribution of model performance metrics. 
The observed trends are consistent across the four target properties: while some architectures generally yield slightly better results, the 
most critical hyperparameter is the learning rate, with optimal performance achieved for values of $10^{-4}$ and $10^{-5}$.

\begin{figure*}
    \centering
    \includegraphics[width=0.9\linewidth]{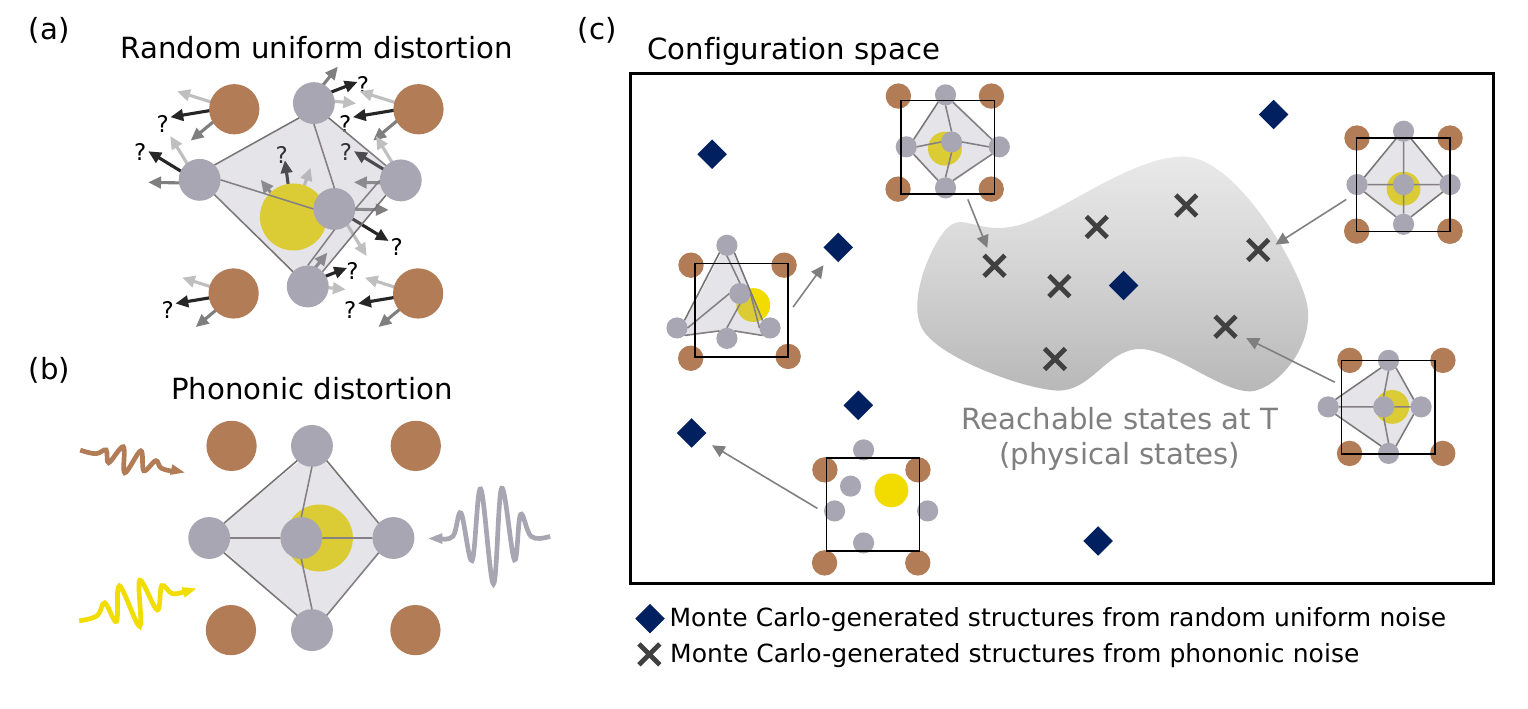}
	\caption{\textbf{Atomically perturbed anti-perovskite structures and configuration space}. Reference anti-perovskite structure 
	perturbed following (a)~random- and (b)~phonon-based atomic displacement schemes. Wavy arrows represent phonon modes, where the 
	color indicates the displaced atom. (c)~Configurational space of possible structures with fixed lattice parameters, where each 
	point corresponds to a specific atomic configuration. The gray region indicates thermally accessible states starting from the 
	reference structure. Diamonds and crosses represent non-equilibrium structures generated by Monte Carlo sampling following 
	random- and phonon-based schemes, respectively.}
    \label{fig3}
\end{figure*}

\subsection{GNN model performance}
\label{subsec:performance}
The dependence of the performance of our GNN models on dataset type and dataset size were thoroughly analysed. For this part of our 
study, we focused on the prediction of the band gap since this quantity exhibits the largest dispersion and lowest $R^2$ among all 
studied cases (Fig.~\ref{fig2} and Table~I). As for the dataset, atomically disordered anti-perovskite configurations were generated 
using two different approaches, described in detail next, one consisting of random uniform displacements and the other being physically 
motivated (Figs.~\ref{fig3}a,b). To ensure good configurational space sampling, all displacements were generated following a Monte 
Carlo procedure (Fig.~\ref{fig3}c). 

Starting with a perfect $40$-atoms anti-perovskite simulation cell, that is, with all the ions located at their equilibrium lattice 
positions, we applied random atomic displacements according to the expression:
\begin{equation}
	u_{j,\alpha}^{\rm rand} = \epsilon \cdot a
\label{eq:random}	
\end{equation}
where $j$ is an atom index, $\alpha$ represents Cartesian direction, $\epsilon$ is a random number sampled from the uniform distribution 
$\epsilon \in \mathcal{U}[-1,1]$, and $a$ is a characteristic displacement length set to $0.8$~\AA~ in this study. This $a$ value was 
chosen to ensure adequate sampling of electronically insulating configurations (Supplementary Fig.~8).

For the physically-informed dataset, the atomic displacements were generated according to the phonon eigenvectors and amplitudes calculated 
for the selected anti-perovskite systems. In particular, we used the well-known expression obtained within the quasi-harmonic approximation 
\cite{togo2023first}:
\begin{eqnarray}
	&	u_{j,\alpha}^{\rm phon}  =  \epsilon \cdot \zeta_{j,\alpha} \nonumber \\
	&  \zeta_{j,\alpha}^{2}  =  \frac{\hbar}{2Nm_{j}} \sum_{\mathbf{q},\nu} \frac{1}{\omega_{\nu}(\mathbf{q})} 
	\left( 1 + 2n_{\nu}(\mathbf{q}, T) \right) \left| e_{\nu}^{\alpha} (j, \mathbf{q}) \right|^{2}~
\label{eq:phon}
\end{eqnarray}
where $\hbar$ is the reduced Planck constant, $N$ is the total number of atoms in the unit cell, $m_{j}$ is the atomic mass of atom $j$, 
$\mathbf{q}$ denotes a reciprocal-space vector, $\nu$ is a phonon mode branch index, and $\omega_{\nu}$, $n_{\nu}$, and $e_{\nu}$ 
correspond to the phonon frequency, Bose-Einstein occupation distribution, and phonon eigenvector, respectively.

\begin{figure*}
    \centering
    \includegraphics[width=1\linewidth]{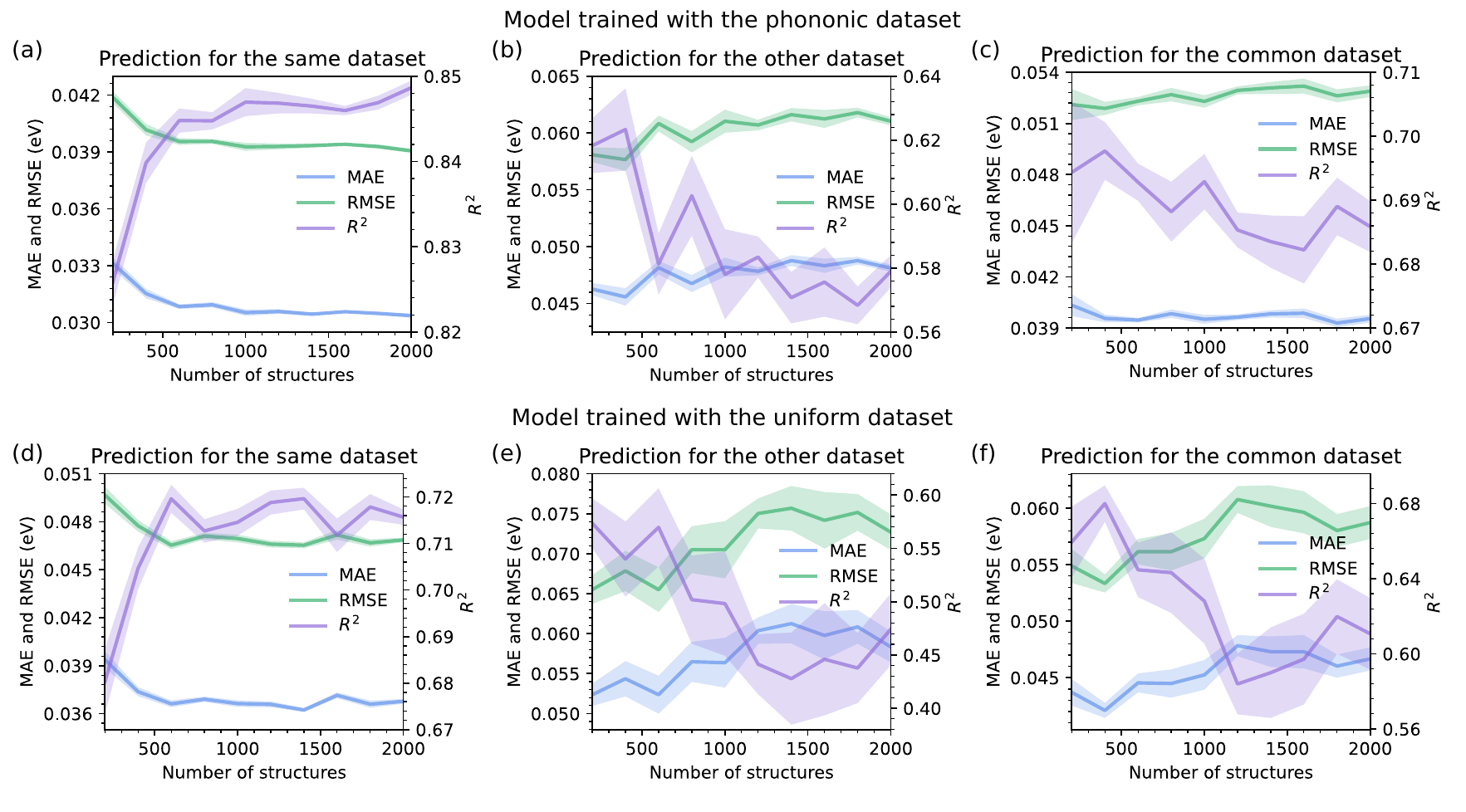}
    \caption{\textbf{Performance of GNN models trained on different datasets for the prediction of band gaps}. Training was performed
	on datasets generated according to (a,b,c)~phonon-based and (d,e,f)~random displacement-based schemes. Models performance tests 
	done on (a,d)~their corresponding dataset, (b,e)~their complementary dataset, and (c,f)~a combined dataset comprising both 
	atomic-displacement generation schemes. Solid lines indicate average values and shaded areas statistical errors.}
    \label{fig4}
\end{figure*}

Two independent band-gap datasets were generated by performing DFT calculations (Methods) on $2,250$ structures obtained separately 
using the random and phonon-informed approaches described above. Independent GNN models were then trained on each dataset, with $250$ 
configurations initially reserved for testing. The convergence of the final GNN models with respect to dataset size was systematically 
monitored, employing multiple random seeds (up to $10$) to ensure statistically meaningful averages and uncertainties. Each model was 
subsequently evaluated on three distinct datasets: (1)~the dataset generated using the same atomic-displacement scheme as in training, 
(2)~the dataset obtained with the complementary displacement approach, and (3)~a combined dataset containing structures generated using 
both the random and phonon-informed methods. The resulting model performance analysis is presented in Fig.~\ref{fig4}.

For both atomic-displacement schemes, the GNN models converge at approximately $1,000$ training structures (Figs.~\ref{fig4}a,d). However, 
the phonon-informed model exhibits superior accuracy, achieving $R^2 \approx 0.85$ and MAE $\approx 0.030$~eV, compared to the 
random-distortion model, which yields $R^2 \approx 0.72$ and MAE $\approx 0.036$~eV. Remarkably, the phonon-based model even outperforms 
the model trained on the combined dataset (Table~I), which includes twice as many structures, indicating that physically consistent data 
is more valuable than simply increasing dataset size. This observation can be rationalized as follows: although random distortions sample 
a broader configurational space, they may include nonphysical atomic arrangements that hinder the model’s ability to learn meaningful 
relationships between bonding geometry and band-gap variations. We will return to this important point, related to the explainability of 
GNN models, in the next section.

\begin{figure*}
    \centering
    \includegraphics[width=1\linewidth]{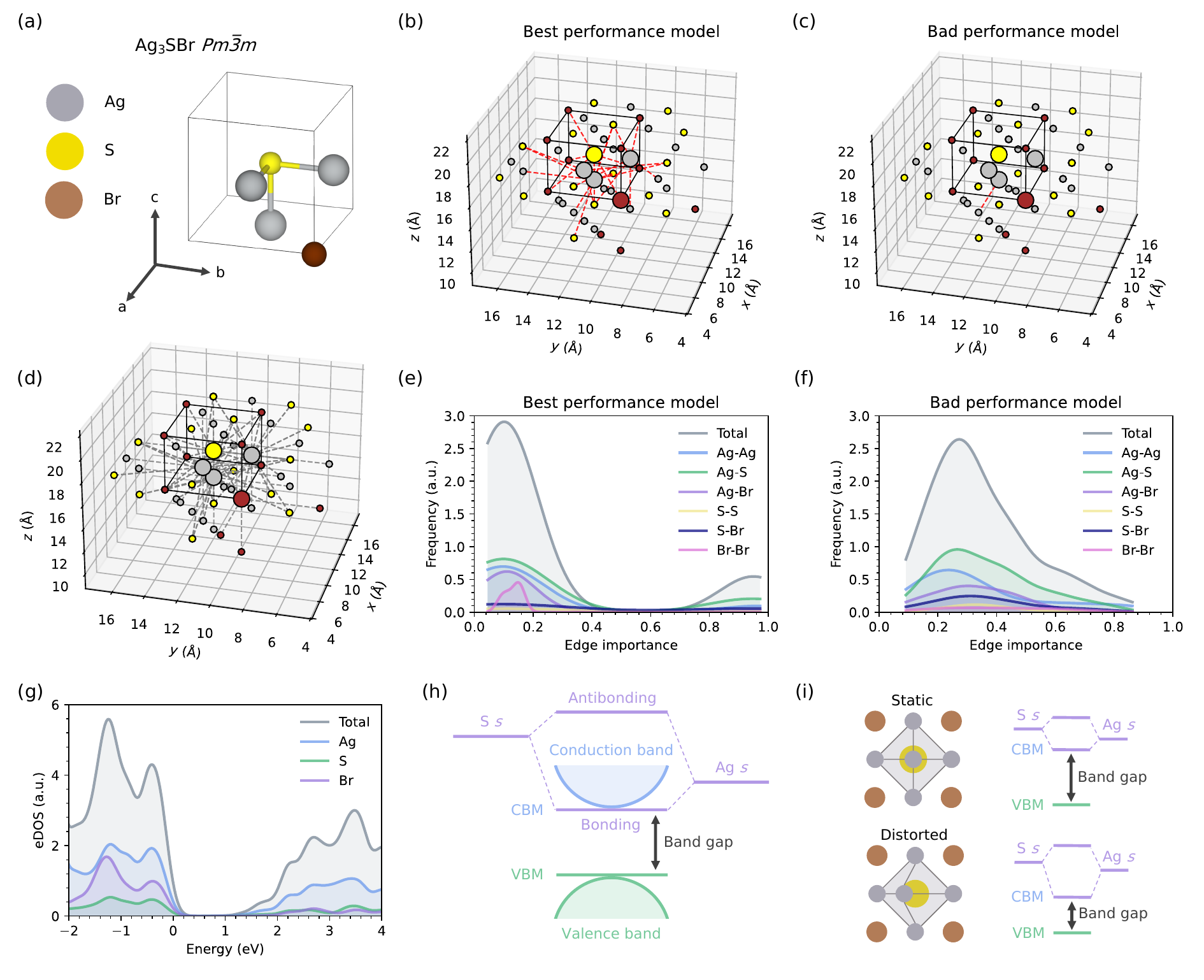}
    \caption{\textbf{GNN model explainability for band gap prediction}. (a)~Reference unit cell for Ag$_3$SBr. (b,c)~Mixed structure–graph 
	representation of the system, where large circles represent unit-cell atoms and small circles their periodic images. Red dashed 
	lines mark edges with importance greater than $0.86$, for (b)~the best-performing GNN and (c)~a poorly performing GNN. The unit 
	cell is highlighted in black. (d)~Graph representation showing all edges as gray dashed lines. (e,f)~Edge-importance density 
	distributions for the best-performing and poorly-performing GNN, compared with the total and pairwise edge densities. (g)~Electronic 
	density of states near the band gap computed with DFT methods. (h)~Electronic structure of Ag$_3$SBr around the band gap, 
	highlighting valence and conduction bands and the orbital hybridization of Ag and S $s$ electrons. (i)~Band-gap reduction 
	resulting from atomic position perturbations in Ag$_3$SBr.}
    \label{fig5}
\end{figure*}

Cross-testing further confirms the superior performance of the GNN model trained on the physics-informed dataset. When evaluated on 
their complementary datasets (Figs.~\ref{fig4}b,e), both models show a decrease in accuracy, consistent with a reduced extrapolation 
capability. Nevertheless, the phonon-informed model generalizes better to the randomly-displaced dataset, as evidenced by its cross-test 
MAE, which is approximately $0.010$~eV lower than that of the model trained on the random dataset. A similar trend is observed for the 
$R^2$ correlation coefficient, which reaches its highest value for the physics-informed model. These results strongly indicate that 
training on physically meaningful data yields more transferable and robust representations.

Finally, both GNN models were tested on a combined dataset containing configurations generated using the two atomic-displacement schemes 
(Figs.~\ref{fig4}c,f). Once again, the phonon-informed model outperforms the random-distortion model, achieving a MAE of $< 0.040$~eV, 
compared with $> 0.045$~eV for the latter. These findings further reinforce the conclusion that data quality and physical relevance 
outweigh dataset size in determining GNN performance for predicting the band gaps of anti-perovskite materials.

It is instructive to analyze the distributions of the four target physical quantities (i.e., atomic energy, band gap, valence band maximum, 
and hydrostatic stress) computed with DFT for the three datasets used to assess GNN model performance. As shown in Supplementary Fig.~9, 
the energy per atom and hydrostatic stress distributions differ substantially among the datasets: the randomly generated dataset contains 
numerous high-energy, nonphysical configurations, whereas the phonon-informed dataset comprises lower-energy, thermally accessible 
structures. These pronounced differences complicate direct comparisons between models trained to predict atomic energies or hydrostatic 
stress on distinct datasets, since their underlying property distributions vary significantly. In contrast, the band-gap distributions are 
nearly identical across all three datasets (Supplementary Fig.~9), enabling a fair and consistent comparison of model performance.

Furthermore, we analyzed the distributions of pairwise atomic distances within the configurations forming the random and phonon-informed 
datasets (Supplementary Fig.~10). Only interatomic distances shorter than $6.0$~\AA~ were considered, as the graph cutoff radius was set 
to $5.5$~\AA, rendering larger distances irrelevant for model construction. For all atomic pairs in the anti-perovskite structure, the 
distance distributions are nearly identical between the two datasets. This structural similarity further ensures that comparisons between 
GNN models trained for band-gap prediction on the two datasets are physically meaningful.

\subsection{GNN model explainability}
\label{subsec:explainability}
To gain insight into why certain GNN models outperform others, we employed graph explainability techniques as implemented in PyTorch Geometric 
\cite{ying2019gnnexplainer,amara2022graphframex}. These methods quantify the relative importance assigned by the GNN to each edge during 
the message-passing process, thus effectively identifying which chemical bonds the model considers most relevant for predicting the band 
gap.

Figure~\ref{fig5}a displays the ideal five-atom cubic anti-perovskite unit cell, providing a structural reference for the subsequent 
analysis. Figures~\ref{fig5}b–d illustrate the relationship between the atomic structure and its corresponding graph representation. 
Large spheres represent the atoms within the unit cell, while smaller ones denote their periodic images. The color scheme differentiates 
atomic species, and the unit cell boundaries (solid lines) are shown for clarity. The most influential edges, as determined by the 
explainability algorithm, are highlighted as red dashed lines for the best-performing model (Fig.~\ref{fig5}b) and for a poorly performing 
model (Fig.~\ref{fig5}c). For reference, all graph edges are shown as gray dashed lines in Fig.~\ref{fig5}d. Edges with an importance 
score above $0.86$ (on a scale from $0$ for least relevant to $1$ for most relevant) were classified as significant. The best-performing 
model identifies several highly important bonds, predominantly Ag--S connections, whereas the low-performing model recognizes only a 
single relevant edge, corresponding to an Ag--Ag bond.

Figures~\ref{fig5}e,f present the edge-importance density distributions for all bonds and for each bond type, comparing the best and 
a poor-performing models. The optimal model assigns low importance (below $0.4$) to most bonds but very high importance (above $0.8$) 
to a few, primarily Ag--S bonds. In contrast, the poorly performing model exhibits a more uniform importance distribution, with most 
bonds receiving moderate scores below $0.6$. This behaviour indicates that the best model effectively distinguishes between chemically 
and electronically relevant bonds and those that play a minor role in determining the band gap, whereas the weaker model fails to make 
such distinctions, effectively treating all bonds as equally relevant.

The prominent emphasis placed on Ag--S bonds by the best-performing model is physically well justified. Previous first-principles studies
\cite{benitez2025giant,benitez2025band} demonstrated that Ag--S interactions are primarily responsible for the temperature dependence of 
the band gap in Ag$_3$SBr. Figure~\ref{fig5}g shows the DFT-calculated electronic density of states (eDOS) for the static paraelectric 
structure, revealing pronounced Ag and S orbital contributions near the conduction band edge. Hybridization between Ag–$s$ and S–$s$ 
orbitals generates bonding and antibonding states, with the conduction band associated with the bonding state (Fig.~\ref{fig5}h). When 
soft phonon modes distort the lattice, the energy splitting between bonding and antibonding states increases, lowering the energy of 
the bonding state, and consequently of the conduction band, resulting in a reduced band gap (Fig.~\ref{fig5}i). Thus, the 
temperature-induced band-gap variation in silver chalcohalide anti-perovskites is predominantly governed by Ag--S bonding.

The ability of the best GNN model to assign high importance to Ag--S bonds therefore demonstrates that it has effectively learned the 
underlying physics controlling the thermal evolution of the band gap. By contrast, Br--Br bonds, which contribute negligibly near the 
band edges, receive very low importance values (below $0.2$). The poorly performing model, however, does not clearly distinguish between 
bond types, assigning similar importance to all of them, evidence that it has not internalized the relevant chemical–electronic 
relationships.

\section{Discussion}
\label{sec:discussion}
The results presented in this work highlight the critical importance of physically informed data generation in the development of ML models for 
materials prediction. By comparing models trained on datasets constructed using random uniform distortions and phonon-informed distortions, we 
demonstrated that data quality and physical relevance often outweigh dataset size. GNN models trained on phonon-informed data achieved superior 
performance in predicting band gaps, even with fewer training structures. This finding underscores that targeted sampling of physically meaningful 
regions of configurational space enhances both generalization and interpretability.

The superior accuracy of the phonon-based models can be rationalized by noting that phonon distortions capture the thermally accessible structural 
configurations of materials (Fig.~\ref{fig3}c). In contrast, random distortions probe unphysical regions of configurational space, including 
unrealistically short interatomic distances ($<0.1$~\AA) or completely disordered structures, which obscure the true structure-property relationships 
the model must learn. These results indicate that incorporating physics through phonons, symmetry constraints, or the curvature of the potential-energy 
surface into data generation guides the model toward the most relevant degrees of freedom, effectively embedding prior physical knowledge within 
the dataset.

Explainability analyses further support this interpretation. The best-performing GNN models correctly identified chemically relevant bonds, 
particularly Ag--S, as the dominant contributors to band-gap evolution under finite-temperature conditions. Poorly performing models, by contrast, 
exhibited diffuse and uncorrelated bond-importance distributions, reflecting a lack of understanding of the underlying bonding physics. This clear 
link between interpretability and predictive accuracy reinforces that explainability tools are essential not only for model transparency but also 
for verifying whether ML models have genuinely learned meaningful structure-property correlations.

These insights are also directly relevant to the development of machine learning interatomic potentials (MLIPs) \cite{jacobs2025practical}. In MLIP 
construction, the quality and representativeness of the training dataset determine both the model’s transferability and its physical reliability. 
Our findings suggest that building training datasets using phonon-based or otherwise physically constrained atomic displacements can yield more 
accurate and stable MLIPs, particularly for describing anharmonic effects, finite-temperature dynamics, and phase transitions.

In light of these findings, the use of atomic configurations extracted from high-quality finite-temperature simulations, such as \textit{ab initio} 
molecular dynamics (AIMD), instead of randomly generated structures, is fully justified and highly recommended for MLIP and GNN model training. 
However, as noted in the Introduction, AIMD simulations are computationally demanding, and their limited sampling times can result in redundant 
exploration of configurational space. In contrast, the dataset generation approach proposed in this study, based on phonon eigenvectors and 
amplitudes, requires minimal computational effort and is straightforward to implement with standard methods. This work therefore provides an 
efficient and physically grounded pathway toward accelerated, interpretable, and reliable ML-driven materials discovery.

\section{Conclusions}
\label{sec:conclusions}
This study demonstrates that incorporating physics into data generation, here through phonon-informed atomic displacements, substantially enhances 
the accuracy, generalization, and interpretability of graph-based ML models for materials prediction. We show that physically meaningful data can 
outperform much larger random datasets, highlighting that dataset design grounded in fundamental principles is as crucial as model architecture. 
Beyond modeling the optoelectronic properties of anti-perovskite systems under realistic $T \neq 0$~K conditions, our findings establish a 
broadly applicable strategy for accelerating and rationalizing ML-driven materials discovery, offering a path toward more efficient and explainable 
models across diverse material classes.

\section*{Methods}
\subsection*{DFT calculations}
DFT calculations \cite{cazorla2017simulation,blochl1994projector} were performed with VASP software 
\cite{kresse1993ab,kresse1996efficiency,kresse1996efficient} and semilocal PBEsol exchange-correlation functional \cite{perdew2008restoring}. 
Wave functions were represented in a plane-wave basis set truncated at $700$~eV. We selected a dense k-point grid, with $12 \times 12 \times 12$ 
points for the reciprocal-space Brillouin zone sampling for the cubic anti-perovskite unit cell. We obtained zero-temperature energies converged 
to within $0.5$~meV per formula unit. For geometry relaxations, a force tolerance of $0.005$~eV~\AA$^{-1}$ was imposed in all the atoms.

\subsection*{Graph generation}
Several methods exist for generating graph representations of crystal structures, including radius cutoffs, $k$-nearest neighbors (k-NN), and 
Voronoi-based approaches \cite{fung2021benchmarking}. In this work, we employed the widely used radius-cutoff method. Each atom in the unit 
cell is represented as a node, characterized by four atomic features: atomic number, atomic mass, atomic radius, and electronegativity. A cutoff 
radius $R_{\mathrm{cutoff}}$ defines the maximum interatomic distance considered for forming edges (i.e., chemical bonds). Two atoms are connected 
by an edge if their Euclidean separation is smaller than $R_{\mathrm{cutoff}}$, and the corresponding distance is stored as the edge feature. To 
ensure periodicity, we build a supercell large enough to fully contain all cutoff spheres centered on each atom. If an atom forms a bond with a 
periodic image of itself, this connection is represented as a self-edge. Further details about the graph construction, including sensitivity tests 
with respect to $R_{\mathrm{cutoff}}$, are provided in the Supplementary Information (Supplementary Figs.~1,2 and Supplementary Discussion). The 
python scripts used for graph generation are openly available in our GitHub repository: \url{https://github.com/polbeni/GNN-materials}. Please 
note that this repository also contains additional scripts and utilities not directly discussed in this work.

\subsection*{GNN models}
Graph Neural Network models were implemented using the PyTorch Geometric framework \cite{fey2019fast}, which is built on top of the PyTorch machine 
learning library \cite{paszke2019pytorch}. Details regarding the specific model architectures, retraining procedures, and hyperparameter optimization 
are provided in the Supplementary Information.

\section*{Data availability}
All relevant scripts, including model training workflows and data preprocessing routines, are freely accessible in the GitHub repository: 
\url{https://github.com/polbeni/GNN-materials}.

\section*{Acknowledgments}
P.B. acknowledges support from the predoctoral program AGAUR-FI ajuts (2024 FI-1 00070) Joan Oró, which is backed by the Secretariat of Universities 
and Research of the Department of Research and Universities of the Generalitat of Catalonia, as well as the European Social Plus Fund. C.L. 
acknowledges support from the Spanish Ministry of Science, Innovation and Universities under an FPU grant. C.C. acknowledges support by 
MICIN/AEI/10.13039/501100011033 and ERDF/EU under the grants PID2023-146623NB-I00 and PID2023-147469NB-C21 and by the Generalitat de Catalunya under 
the grants 2021SGR-00343, 2021SGR-01519 and 2021SGR-01411. Computational support was provided by the Red Española de Supercomputación under the 
grants FI-2024-1-0005, FI-2024-2-0003, FI-2024-3-0004,FI-2024-1-0025, FI-2024-2-0006, and FI-2025-1-0015. This work is part of the Maria de Maeztu 
Units of ExcellenceProgramme CEX2023-001300-M funded by MCIN/AEI (10.13039/501100011033). E.S. acknowledges the European Union H2020 Framework Program 
SENSATE project: Low dimensional semiconductors for optically tuneable solar harvesters (grantagreement Number 866018), Renew-PV European COST action 
(CA21148) and the Spanish Ministry of Science and Innovation ACT-FAST (PCI2023-145971-2). E.S. is grateful to the ICREA Academia program.

%\bibliographystyle{unsrt}
%\bibliography{bibliography.bib}

\end{document}